\documentclass[final,1p,times]{elsarticle} 
\usepackage{graphicx} 
\usepackage{amssymb} 
\usepackage{amsthm} 
\usepackage{lineno} 

\usepackage{amsmath}

\newcommand{\GEV} {GeV/{\it c}}
\newcommand{\benergy}{$\sqrt{s_{NN}}=200$~GeV}
\newcommand{\pp} {{\it p\,+\,p}}
\newcommand{\raajet}{$R_{AA}^\textrm{jet}$~}
\newcommand{\figWH}{4.9cm}

\journal{Nuclear Physics A} 
\begin{document} 

\begin{frontmatter} 


\title{Inclusive cross section and correlations of fully reconstructed jets in $\sqrt{s_{NN}}=200$~GeV~Au\,+\,Au and \pp\ collisions}

\author{Mateusz P\l osko\'n for the STAR Collaboration}

\address{Lawrence Berkeley National Laboratory\\
1 Cyclotron Road, Berkeley, CA 94720, USA}

\begin{abstract} 
We present an experimental study of full jet reconstruction in 
the high multiplicity environment of heavy ion collisions, utilizing
\benergy~\pp~and central Au\,+\,Au data measured by STAR. Inclusive differential jet production cross
sections and ratios are reported, as well as high-$p_{T}$ hadron--jet coincidences.
\end{abstract} 

\end{frontmatter} 


\section{Introduction}
\label{lab:introduction}
Measurements of jet quenching via single and di-hadron observables
\cite{bib:jet_suppr} have provided initial estimates of the energy
density of the hot QCD medium generated in high energy nuclear
collisions. However, such observables suffer from well-known biases
and are limited in their sensitivity \cite{bib:highpt_surfbias}. Fully
reconstructed jets will enable an unbiased exploration of quenching,
including new observables of energy flow within jets whose theoretical
description is not dependent upon the modeling of hadronization.

Full jet reconstruction is difficult in the heavy-ion collision
environment. Attempts to suppress background via seeded
algorithms or the introduction of a modest cut on hadron $p_T$
in the jet reconstruction result in a jet population biased
against quenched jets \cite{bib:hp08}. Our emphasis in this analysis
is to minimize such biases. We apply the minimum cuts
possible within the STAR acceptance ($p_T > 0.2$~\GEV) and use seedless algorithms in both \pp~and Au\,+\,Au
collisions. We confront the challenging problem of high-density
backgrounds by applying recently developed jet finding and background correction
algorithms of the FastJet package \cite{bib:fastjet}. We report inclusive jet cross-sections in \benergy\ \pp~and
10\% most central Au\,+\,Au collision, using two resolution
scales ($R=0.2$ and $R=0.4$). We report their ratios, jet $R_{AA}$,
and hadron--jet correlations.

The results presented in these proceedings differ quantitatively in
two respects from those presented at the conference, though all
qualitative conclusions are unchanged. For \pp~collisions, full
correction for the jet trigger bias has now been applied (whereas an
approximation was used previously), resulting in an increase in the
central value of the cross section for $p_{T}^{\textrm jet}<30$~\GEV~and
corresponding decrease in $R_{AA}$. For Au\,+\,Au collisions, the study of simulated (Pythia) 
jets embedded in Au\,+\,Au background events resulted in a
revision of the parameterization of background fluctuations. We correct the overall scaling error, which for narrow jets ($R=0.2$) results in a corresponding increase of up to a factor of 3 in the inclusive cross section.

\section{Data sets, algorithms, and procedures}
\label{lab:algors}

For \pp~collisions, we use 12~pb$^{-1}$
recorded in Run 6 with a jet patch trigger, requiring a transverse energy deposition above $7.6$~GeV
in a fixed BEMC region of dimensions $\Delta \eta \times
\Delta \phi = 1 \times 1~{\textrm rad}$. For central Au\,+\,Au collisions, we use the 7.6 $\times~10^6$ most central
events ($\sigma /\sigma_\textrm{Geom} = 10\%$) recorded in Run 7 with
a minimum bias trigger. Jet reconstruction in both heavy-ion and
\pp~collisions utilizes the $k_{T}$ \cite{bib:kt} and anti-$k_{T}$ \cite{bib:antikt} 
sequential recombination algorithms provided by the FastJet
package \cite{bib:fastjet}. The two algorithms have different sensitivities to
heavy-ion background. The energy recombination scheme is used for
tracks and calorimeter towers, both of which are assigned zero mass. Jet
candidates are found utilizing two resolution parameters $R=0.2$ and
$R=0.4$, within acceptance $|\eta^\textrm{det}| < 0.6$. Fragmentation
biases are minimized through the minimal cuts
($p_T^\textrm{track/tower} > 0.2$~\GEV~and
$|\eta^\textrm{track/tower}| < 1$). BEMC towers matched with one
or more TPC tracks are removed from analysis.

Background subtraction is applied event-wise via $p_{T}^{\textrm{rec}}
= p_{T}^{\textrm{candidate}} - \rho \cdot A$, where $\rho$ measures
$p_T$-weighted density of background in an event and $A$ is the
measured jet area \cite{bib:fastjet}. We define signal jets with
$p_T^{\textrm{rec}} > 0$ and background jets with $p_T^{\textrm{rec}}
\leq 0$. For central Au\,+\,Au collisions, $\rho \approx 75$~\GEV~per unit
area within $-0.6<\eta <0.6$. The term $\rho \cdot A$ is the most
probable value of the background underlying the signal
jet. Fluctuations around this value are approximated by a Gaussian
distribution with width of $\sigma^{\textrm{bg}}$, which is applied in
an unfolding procedure to correct the inclusive spectrum. The value of
$\sigma^{\textrm{bg}}$ extracted from the distribution of
background jets reproduces the spectrum distortion of Pythia jets
embedded into central Au\,+\,Au events only for $R=0.4$, but not for
$R=0.2$. This may be related to the non-Gaussian nature of the
background fluctuations; the Pythia embedding is more sensitive to the
upwards fluctuations, which are predominantly responsible for the
spectrum distortion. We therefore utilize in these proceedings the
correction factors extracted from embedded Pythia using 
$\sigma^{\textrm{bg}}_{R=0.4} = 6.8$~GeV and
$\sigma^{\textrm{bg}}_{R=0.2} = 3.7$~GeV. This choice results in an
increase of a factor $\approx 3$ in the inclusive jet cross section for
$R=0.2$. Prior to the unfolding procedure, the signal jet spectrum is corrected
for ``false''-jet yield, defined as the signal in excess of the
background model from random association of uncorrelated soft
particles, estimated by running the jet finders on a
randomized Au\,+\,Au event, with jet-leading particles removed.

The jet energy correction for tracking inefficiency, approximated in this analysis to be
$p_T^{\textrm{track}}$-independent, is 8\% in \pp~ and~12\% in Au\,+\,Au data, 
and correction for unobserved neutral energy (such as neutrons and $K_0^L$) is 5\% in both systems. The total instrumental jet energy
resolution ($\sigma^\textrm{det}(E)/E \approx 20\%$ \cite{bib:Helen}) is corrected
by unfolding. The systematic uncertainty on the jet energy scale,
shown in all figures as the vertical shaded band, is dominated by the
uncertainties on the BEMC calibration (5\%), unobserved jet energy (3\%), and charged
component momentum resolution (2\%). Solid lines represent systematic
uncertainty on the jet yield in Au\,+\,Au due to background fluctuations.
\section{Inclusive cross-sections and ratios}
\label{sec:jetxsections}
\label{lab:xsections}
\begin{figure}[htbp]
	\begin{minipage}{10cm}
		\centering
		\includegraphics[width=\figWH, height=\figWH]{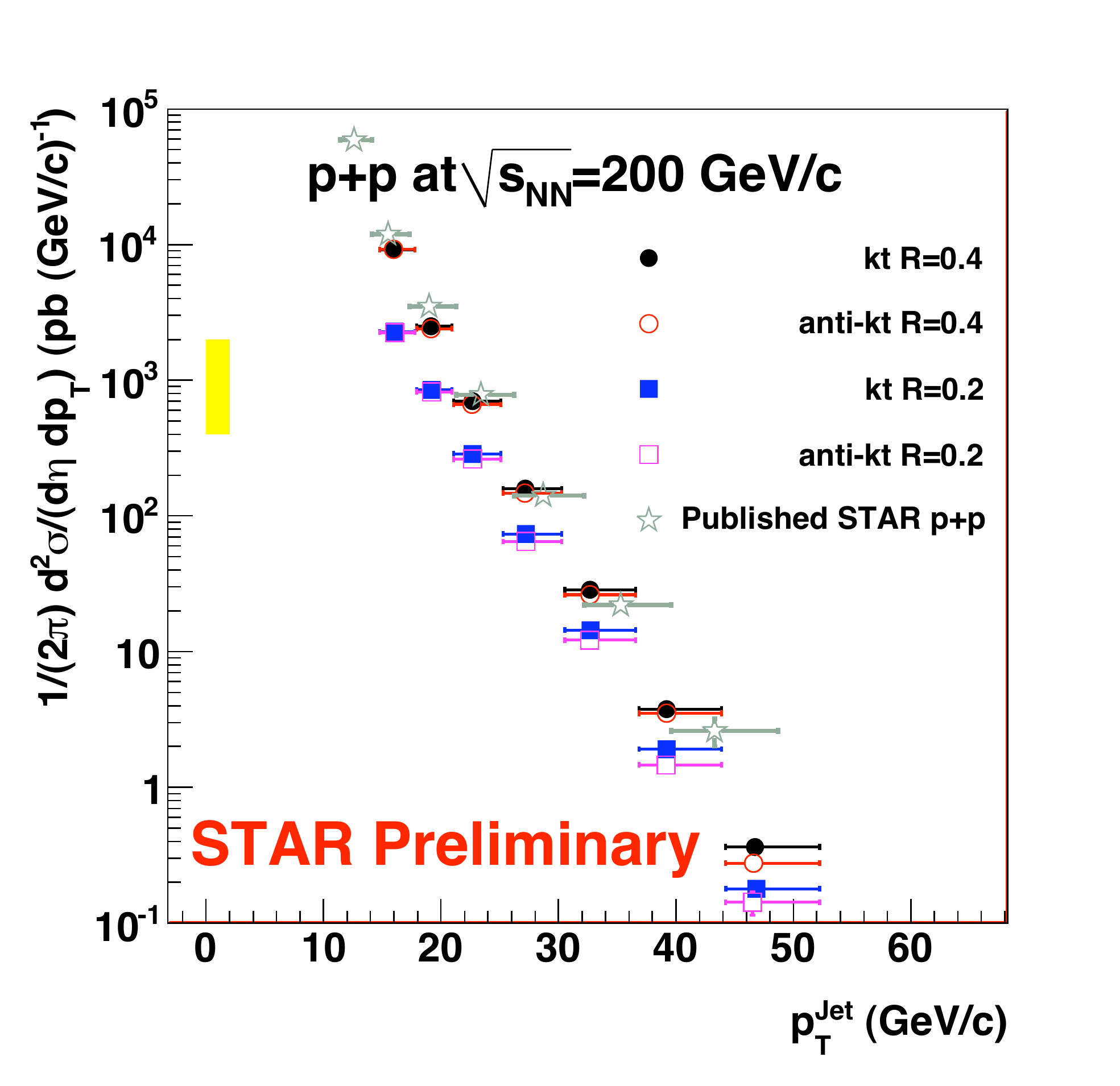} 
		\includegraphics[width=\figWH, height=\figWH]{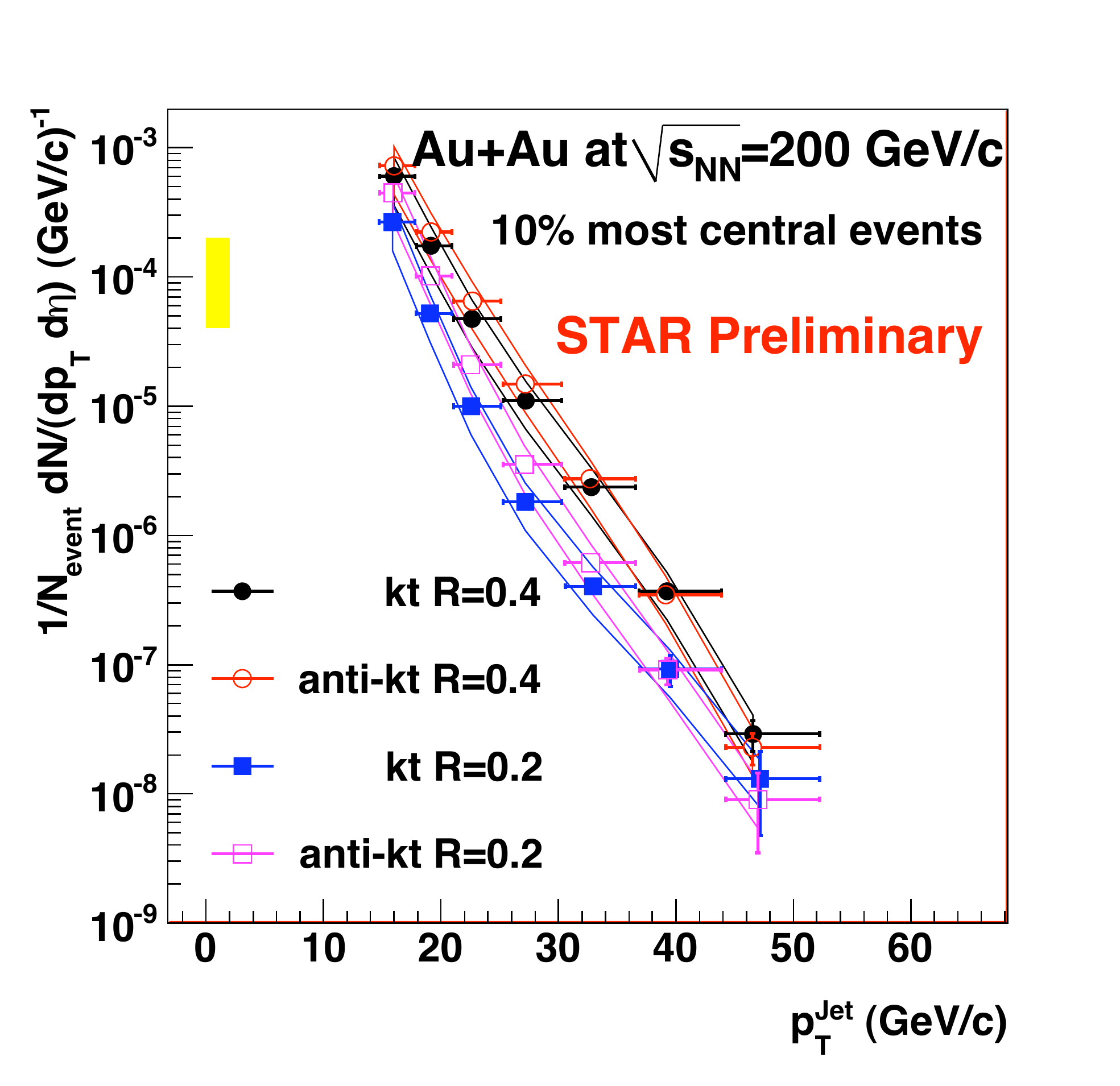} 
 	\end{minipage}
	\begin{minipage}{3cm}
	\caption{ Cross sections for inclusive jet production in~\pp~({\it left}) and Au\,+\,Au collisions ({\it right}) at $\sqrt{s_{NN}}=200~$GeV ($k_{T}$ and anti-$k_{T}$, {\it R} = 0.2 and 0.4). Error bands described in the text. Published \pp~data are from Ref.~\cite{bib:star_jetpaper}.}		
	\end{minipage}
\label{fig:xsections}
\end{figure}
Figure \ref{fig:xsections} presents the differential cross sections for
fully reconstructed inclusive jet production in \pp~and in 10\% most central Au\,+\,Au
collisions at \benergy. The reconstructed jet yields extend in
$p_T^\textrm{jet}$ beyond 50~\GEV. The \pp~measurement agrees within uncertainties with the 
published cross section (mid-point cone, $R=0.4$) \cite{bib:star_jetpaper}.

\begin{figure}[htbp]
	\begin{minipage}{10cm}
		\centering
 		\includegraphics[width=\figWH,height=\figWH]{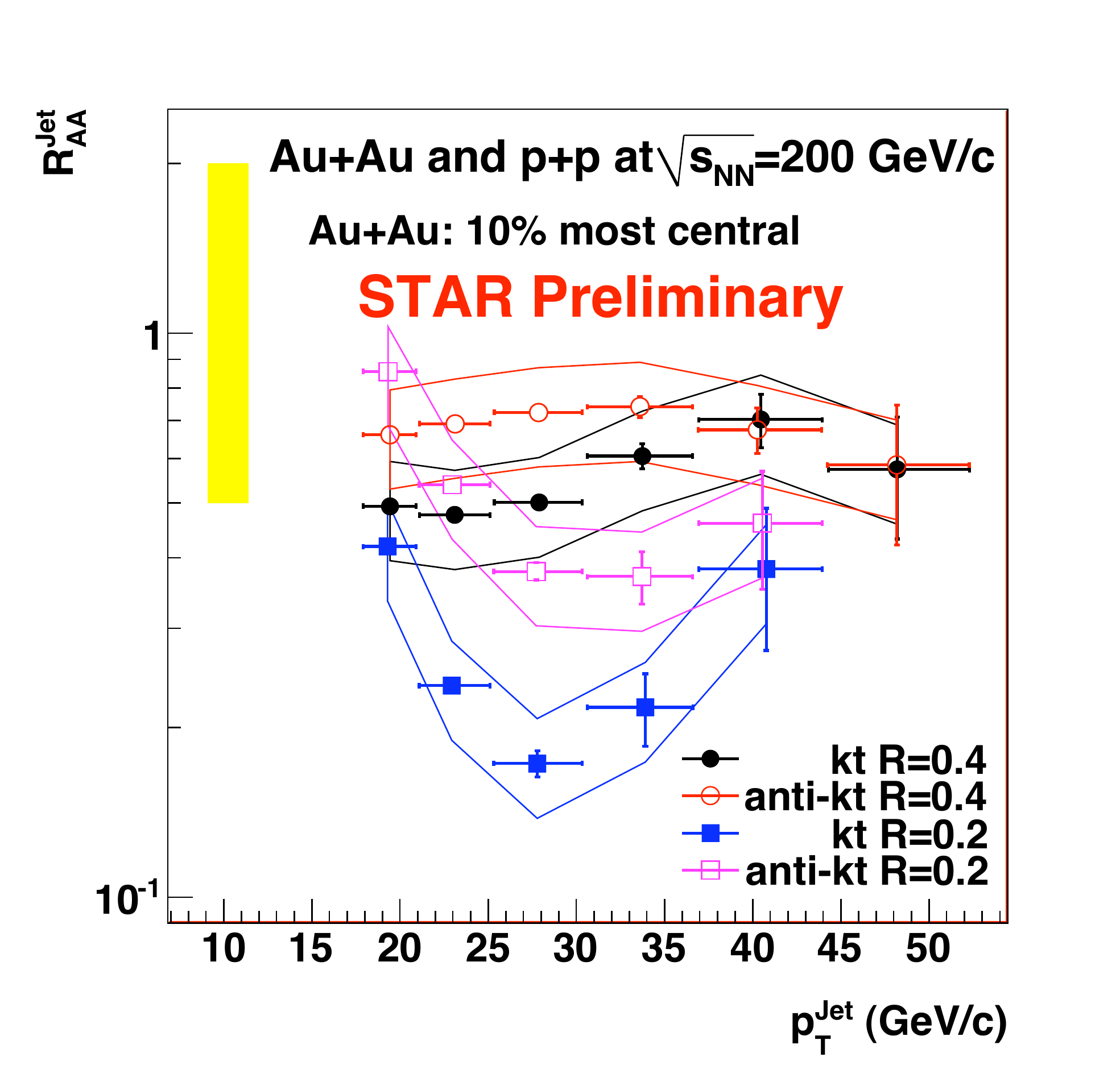} 
 		\includegraphics[width=\figWH,height=\figWH]{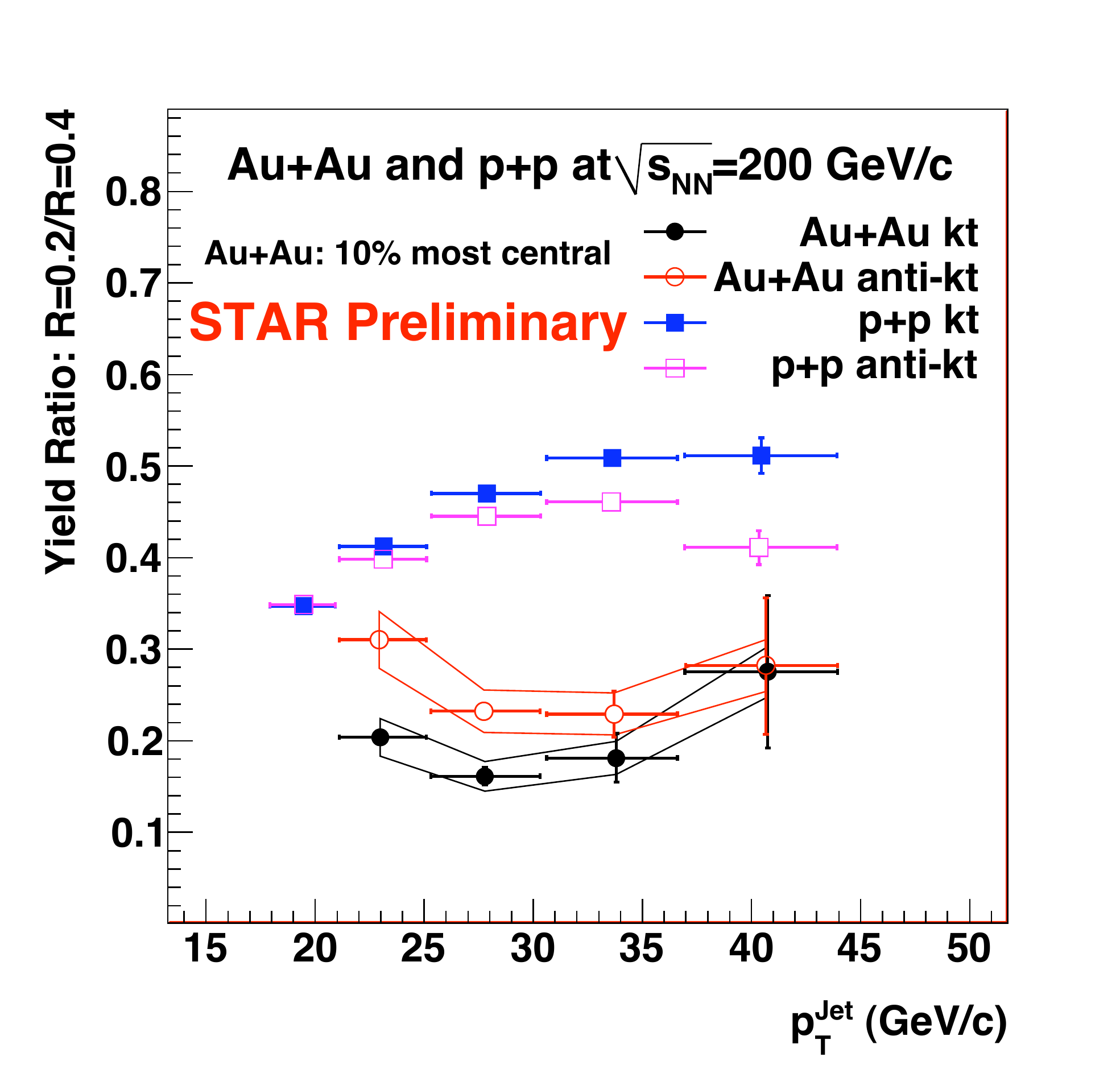}
	\end{minipage}
	\begin{minipage}{3cm}
		\caption{Ratios of inclusive jet
 		cross sections in \pp~and Au\,+\,Au collisions at \benergy~ ($k_{T}$ and
 		anti-$k_{T}$, {\it R} = 0.2 and 0.4).  {\it Left: } Jet $R_{AA}$. {\it
 		Right: } Ratio of cross sections $R=0.2/R=0.4$ for each system.}
	\end{minipage}
\label{fig:ratiosxsections} 
\end{figure}

Figure \ref{fig:ratiosxsections} (left panel) shows \raajet, the ratio
of the jet yield in Au\,+\,Au over the binary collision-scaled jet yield in
\pp.  For unbiased jet reconstruction, this ratio is expected to be
close to unity, with possible deviations due to initial state
effects. We find \raajet for $R=0.4$ compatible with unity, within the large
uncertainties. The \raajet for $R=0.4$ is significantly larger than
$R_{AA}$ of hadrons for $p_{T}<$ 20~\GEV~($R_{AA}^\textrm{hadron}\approx
0.2$). The \raajet for $R=0.2$ is markedly below that for
$R=0.4$. There are significant differences between $k_{T}$ and
anti-$k_{T}$ algorithms, possibly arising from their different response to the
heavy-ion background.

Figure~\ref{fig:ratiosxsections} (right panel) shows the ratio of jet
yield for $R=0.2$ over that for $R=0.4$,
separately for \pp~and Au\,+\,Au collisions. Several jet energy scale
systematic uncertainties cancel in this ratio. For \pp~collisions, the
ratio increases with $p_T^{\textrm{jet}}$, consistent with a Pythia calculation but not with a recent
NLO calculation \cite{bib:WVogelsang}. The ratio is strongly
suppressed for central Au\,+\,Au relative to \pp~collisions, indicating
substantial broadening of the jets in heavy-ion collisions. A recent NLO calculation has been carried out that addressed these measurements directly \cite{Vitev:2009rd}.
\section{Hadron--jet coincidences}
\begin{figure}[htbp]
\begin{minipage}{0.5\textwidth}
\centering
\includegraphics[width=\figWH, height=\figWH]{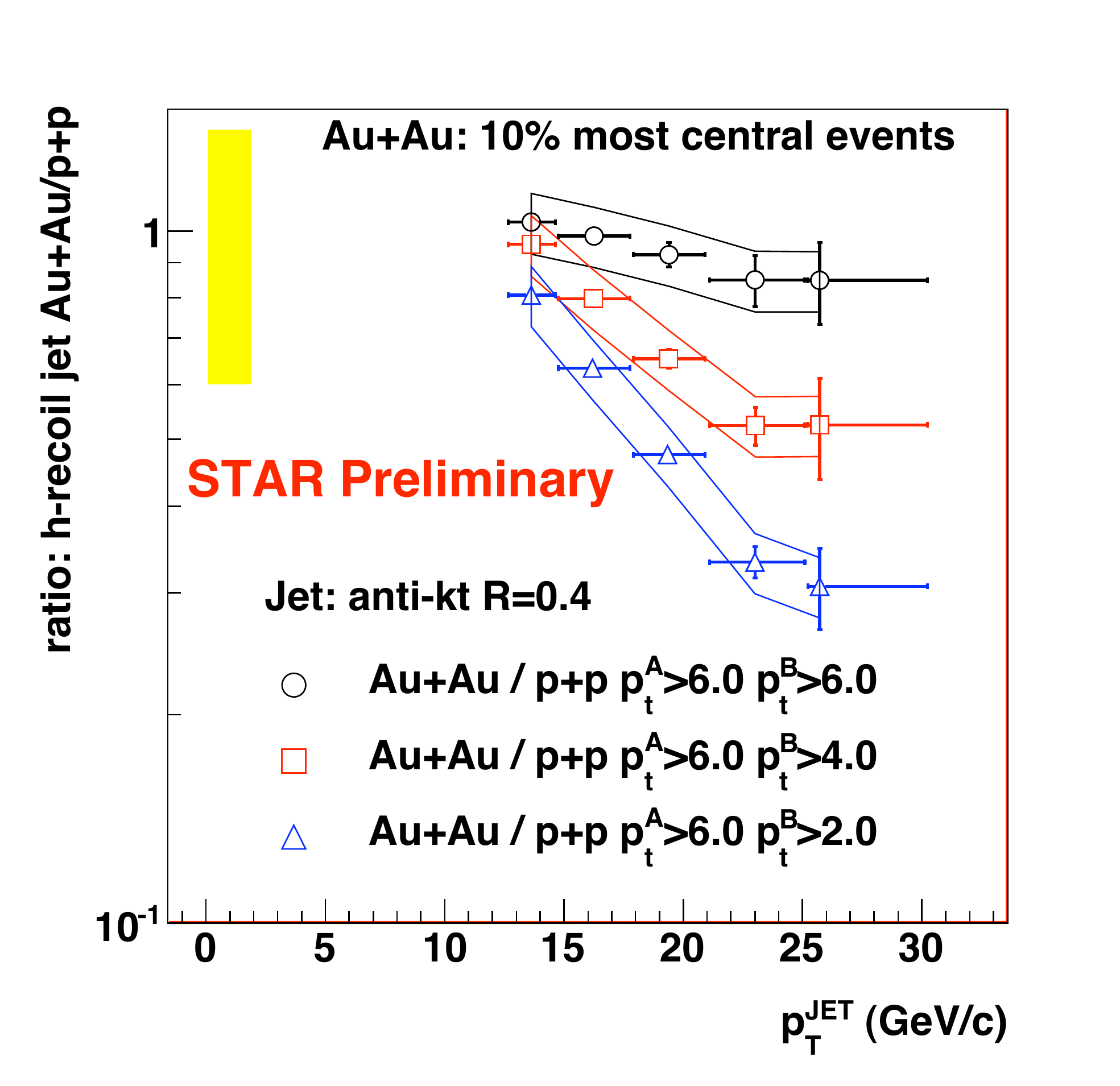} 
\end{minipage}
\begin{minipage}{0.5\textwidth}
\caption{Ratios of hadron--jet conditional yields for three selections of the leading particle of the recoil jet, reconstructed in 10\% most central Au\,+\,Au collisions and in \pp~collisions. Jets were reconstructed with anti-$k_{T}$ algorithm with $R=0.4$ and correlated at $\Delta\phi \approx \pi$ with a leading high tower trigger cluster (dominated by $\pi^{0}$).}
\label{fig:hjet}
\end{minipage}
\end{figure}
We study the correlation of high-$p_{T}$ trigger particles (BEMC clusters
with $p_T > 6$~\GEV) with a recoiling jet (matched in azimuth within
$|\Delta\phi - \pi| < 0.4$), comparing central Au\,+\,Au and \pp~collisions. In
Au\,+\,Au, this exploits the geometric bias of high-$p_{T}$ hadron production
\cite{bib:highpt_surfbias} due to quenching, which maximizes the path
length of the recoiling jet in matter. Additional geometric bias can
be introduced by applying the $p_{T}$ thresholds on the leading hadron in
the recoil jet. The recoil-jet spectra have been corrected for the
``false''-jet contamination, described above. Additional uncorrected
background due to multiple hard interactions in one Au\,+\,Au
collision may be present. The yields are normalized to the number of trigger hadron pairs.

Figure \ref{fig:hjet} shows the ratio of conditional jet yields
(Au\,+\,Au\,/\,\pp) vs. $p_{T}^{jet}$, for various recoil jet leading particle
thresholds. The strongly exclusive requirement of two back-to-back
high-$p_{T}$ particles in Au\,+\,Au biases this measurement towards jets with little
interaction, whereas a more relaxed condition permits more interacting
jets. The ratio observed for the high-$p_{T}$ selection is consistent with
unity, indicating that in Au\,+\,Au collisions we reconstruct the same jet
population per di-hadron trigger as in \pp~collisions. For the low-$p_{T}$
selection, on the contrary, the ratio is below unity and drops significantly
with increasing $p_T^{\textrm jet}$, indicating substantial
redistribution of jet energy in heavy-ion collisions. Related di-jet
correlation studies are presented in Ref. \cite{bib:Elena}.

\section{Summary}
We have presented the cross sections for fully reconstructed
inclusive jet production in \pp\ and Au\,+\,Au collisions at \benergy, obtained with two different jet
algorithms and with two different resolution parameters $R=0.2$ and
$R=0.4$. While \raajet~for $R=0.4$ is compatible with unity
within large uncertainties, more discriminating measurements of the
inclusive cross section as a function of resolution parameter and
hadron--jet correlations indicate strong broadening of the jets 
in heavy-ion collisions.

\end{document}